\documentclass[12pt,a4paper]{article}
\usepackage{graphicx}
\usepackage{amsmath}
\usepackage{amssymb}
\usepackage{epstopdf}
\usepackage{pdfsync}
\usepackage[latin1]{inputenc}
\newcommand{\du}{\mathrm{d}}
\providecommand*{\iu}{\text{i}}
\newcommand{\eg}{{\it e.g.\ }}
\renewcommand{\vec}[1]{
{\boldsymbol#1}}

\title{An accurate boundary value problem solver applied to scattering
  from cylinders with corners}
  \author{%
{Johan Helsing{\small $~^{1}$}, Anders Karlsson{\small $~^{2}$} }%
\vspace{1.6mm}\\
\fontsize{10}{10}\selectfont\itshape
$^{1}$\,Centre for Mathematical Sciences,\\ \fontsize{10}{10}\selectfont\rmfamily\itshape
$^{2}$\,Electrical and Information Technology,\\\fontsize{10}{10}\selectfont\itshape Lund
  University, Box 118, 221 00 Lund, Sweden\\
\fontsize{9}{9}\selectfont\ttfamily\upshape
$^{1}$\,helsing@maths.lth.se, $^{2}$\,anders.karlsson@eit.lth.se}
\begin{document}
\maketitle
\begin{abstract}
  In this paper we consider the classic problems of scattering of
  waves from perfectly conducting cylinders with piecewise smooth
  boundaries. The scattering problems are formulated as integral
  equations and solved using a Nyström scheme where the corners of the
  cylinders are efficiently handled by a method referred to as
  Recursively Compressed Inverse Preconditioning (RCIP). This method
  has been very successful in treating static problems in non-smooth
  domains and the present paper shows that it works equally well for
  the Helmholtz equation. In the numerical examples we specialize to
  scattering of E- and H-waves from a cylinder with one corner. Even
  at a size $kd=1000$, where $k$ is the wavenumber and $d$ the
  diameter, the scheme produces at least 13 digits of accuracy in the
  electric and magnetic fields everywhere outside the cylinder.
\end{abstract}

\section{Introduction}

The numerical simulation of scattering from cylinders has a long
history in computational electromagnetics. As early as in 1881, Lord
Rayleigh treated the scattering of light from a circular dielectric
cylinder~\cite{Rayleigh1881}. He considered an incident plane E-wave,
i.e.,~the electric field is parallel to the cylinder, and a
permittivity and permeability of the cylinder that departed only
slightly from those of the surrounding medium. That enabled him to
find an approximate solution that today is referred to as the Born
approximation and can be viewed as spectral method solution with only
one basis function, c.f.~\cite[Section 8.3.4]{VanBladel1985}. The
theory of scattering from circular cylinders and spheres, conducting
or dielectric, was soon after that fully understood by using
expansions of the incident and scattered waves in partial waves,
c.f.~\cite{Mie1908}. Since then, a large number of papers have been
published that solve scattering problems in electromagnetics, as well
as in acoustics and elastodynamics, using different numerical
techniques. All with the common goal of constructing faster and more
accurate solvers for ever more detailed and complex geometries in two
and three space dimensions. In particular, integral equation methods
have become very important tools. In electromagnetics such methods
were made popular by the contributions of Harrington,
c.f.~\cite{Harrington1968}. The mathematical foundations of the
scattering problems and the integral equation formulations are
discussed in the books by Colton and Kress
\cite{Colton+Kress1983,Colton+Kress1992}.

The present paper is about scattering from piecewise smooth perfectly
conducting objects. The presence of boundary singularities, such as
corners, tends to cause complicated asymptotics in quantities used to
represent the solution. Intense mesh refinement might be needed for
resolution, but this is costly and can easily lead to instabilities
and the loss of precision in the computed field. In the context of
integral equation solvers, regions close to the boundary are the most
problematic. On the application side, scattering from non-smooth metal
objects is of great importance in radar imaging of objects with sharp
corners such as airplanes, vessels and vehicles. Sharp corners that
are oriented perpendicular to the line of sight of a monostatic radar
may create reflections that are large enough to be detected by the
radar. The two-dimensional approximations can be used for elongated
objects like wings but also in the evaluation of fields in the near
zone of smaller objects. Other important two-dimensional problems are
wave propagation in rectangular waveguides, photonic band gap
structures, and substrate integrated waveguides.

The numerical solver used in this paper takes its starting point in a
Fredholm second kind integral equation with integral operators that
are compact away from boundary singularities and whose unknown
quantity is a layer density representing the solution to the original
problem. The integral equation is discretized using a Nystr{\"o}m
scheme and composite Gauss--Legendre quadrature. At the heart of the
solver lies a method called Recursively Compressed Inverse
Preconditioning (RCIP). It modifies the kernels of the integral
operators so that the layer density becomes piecewise smooth and
simple to resolve by polynomials. Loosely speaking one can say that
RCIP makes it possible to solve elliptic boundary value problems in
piecewise smooth domains as cheaply and accurately as they can be
solved in smooth domains. The RCIP method originated in
2008~\cite{Helsing+Ojala2008}. In a series of papers it has been
extended and successfully applied to electrostatic and elastostatic
problems which, at first glance, might seem outright impossible. For
example, the effective conductivity of a high-contrast conducting
checkerboard with a million randomly placed squares in the unit cell
was computed on a regular workstation with a relative accuracy of
$10^{-9}$~\cite{HelsingJCP2011}. A new record has been established for
the three-dimensional problem of determining the capacitance of the
unit cube -- 13 digits compared to the seven digits that were
previously known~\cite{Helsing+Perfekt2012}.

When we here apply the RCIP method to the Helmholtz equation and the
exterior Dirichlet and Neumann problems we do this in a
two-dimensional setting. We consider scattering of time-harmonic E-
and H-waves from an infinitely long perfectly conducting cylinder.
Scattering problems are harder to solve than electrostatic problems,
all other things held equal. Planar problems provide a good testing
ground prior to a move up to three dimensions~\cite{Kloeckner2012}. As
we shall see, the transition from Laplace's equation to the Helmholtz
equation is surprisingly straightforward and the results, presented in
Section~\ref{sec:examples} below, are as good as the ones obtained for
electrostatics.

Our numerical solver meets five important criteria. The first criteria
is that it can handle cylinders with general shapes. In practice this
means cylinders with piecewise smooth boundaries and with a finite,
but arbitrary, number of corners. The second criteria is that it can
treat frequencies ranging from zero up to large values of $kd$, where
$k$ is the wavenumber and $d$ the diameter of the object. We have
found that $kd=1000$ is quite easy to reach and for most cylinders
this frequency range overlaps the frequency band where approximate
high frequency methods, \eg unified theory of diffraction in
combination with physical optics, can be applied with reasonable accuracy.
The third criteria is that the method can deliver accurate results for
the scattered field everywhere outside the object. Even close to a
corner and at $kd=1000$ the scattered field is calculated with at
least 13 digits of accuracy in IEEE double precision arithmetic (16
digit precision). The fourth criteria is that the method enables fast
solvers. In the present implementation the solver is fast only in the
sense that the cost for modifying the kernels of the integral
operators grows linearly with the number of corners in the
computational domain. The method can be made fast {\it in toto} by
incorporating fast multipole techniques~\cite{Carrier1988,Cheng2006}
or perhaps even fast direct solvers~\cite{Martinsson2005,Bremer2012}.
The fifth criteria is that the method is automatized and flexible. It
requires only a minimum of adjustments as operators and geometries
change. 

It is beyond the scope of the present paper to review the RCIP method
in its entirety. In Section~\ref{sec:scheme} we give a brief overview
and a few details on discretization issues particular to Hankel
kernels. Apart from that, we refer readers to the original research
papers~\cite{Helsing+Ojala2008,Helsing+Ojala2009,HelsingJCP2009,HelsingSIAM2011}
and to a newly written tutorial~\cite{Tutorial2012}.

There are several recent journal papers that focus on speed and
accuracy for two-dimensional scattering problems in complex
geometries. In \cite{Tsalamengas2010} scattering from two-dimensional
smooth strips are treated using integral equations and a Nystr{\"o}m
method. In \cite{Tsalamengas+Nanakos2012} the approach of
\cite{Tsalamengas2010} is generalized to smooth slotted cylinders. A
similar problem is treated in \cite{Tsong+Chew2006}. The schemes used
in these papers give accurate results but they cannot, in a simple
way, be generalized to non-smooth geometries. In \cite{Bremer2012} and
in \cite{Greengard2012}, on the other hand, very fast and also
flexible and accurate numerical schemes are developed for the solution
of integral equations modeling scattering from general objects with
both corners and multi-material junctions. These papers, however, do
not address the problem of accurate near field evaluation.

\section{Formulation of the problems}
\label{sec:form}

We consider in-plane waves scattered by a bounded perfectly conducting
cylinder with a piecewise smooth boundary $\Gamma$. The region outside
the object is denoted $\Omega_\text{ex}$, the time dependence is
$e^{-\iu\omega t}$ and $\vec r=(x,y)$. Both E-waves, often referred to
as {T}M-waves, and H-waves, often referred to as TE-waves, are
treated. We decompose the electric and magnetic fields into a sum of
the incident field, denoted $U_\text{inc}(\vec r)$, generated by a
source in $\Omega_\text{ex}$, and the scattered field, denoted
$U_\text{sca}(\vec r)$ in both cases.

\subsection{E-waves}

We let the electric field be parallel to the cylinder, $\vec E(\vec
r)=\hat{\vec z}U(\vec r)$, and let $U(\vec r)=U_\text{inc}(\vec
r)+U_\text{sca}(\vec r)$. The scattered field $U_\text{sca}(\vec r)$
satisfies the following exterior Dirichlet problem:
\begin{align}
&\nabla^2U_\text{sca}(\vec r)+k^2U_\text{sca}(\vec r)=0,\, \vec r\in \Omega_\text{ex}\label{eq1}\\
&U_\text{sca}(\vec r)=-U_{\text{inc}}(\vec r),\, \vec r\in \Gamma\label{eq2}\\
&\lim_{\vert\vec r\vert\rightarrow \infty}\left(\dfrac{\partial}{\partial r}-\iu k\right)U_\text{sca}(\vec r)=0.
\end{align}

We write the solution as the combined integral representation
\cite[eq.~(3.25)]{Colton+Kress1992}.
\begin{equation} \label{representationE}
U_\text{sca}(\vec r)=\int_\Gamma \dfrac{\partial \Phi_k(\vec r,\vec r')}{\partial\nu_{r'}}\rho(\vec r')\du\ell'-\iu\dfrac{k}{2}\int_\Gamma \Phi_k(\vec r,\vec r')\rho(\vec r')\du\ell',\,\vec r\in \Omega_\text{ex},
\end{equation}
where $\Phi_k(\vec r,\vec r')=\dfrac{\iu}{4}H_0^{(1)}(k\vert\vec
r-\vec r'\vert)$ is the free space Green function for the Helmholz
equation in two dimensions, $H_0^{(1)}$ is the Hankel function of the
first kind of order zero, and ${\rm d}\ell$ is an element of arc
length. The index $k$ indicates that the quantity or function depends
on the wavenumber $k=\omega/c$.  Insertion of~\eqref{representationE}
into \eqref{eq2} gives the integral equation for the layer density
$\rho(\vec r)$
\begin{equation}
(I+K_k-\iu \dfrac{k}{2}S_k)\rho(\vec r)=-2U_{\text{inc}}(\vec r), \,\vec r\in \Gamma,
\label{eq:Eint}
\end{equation}
where
\begin{align}
  K_k\rho(\vec r)=2\int_\Gamma \dfrac{\partial \Phi_k(\vec r,\vec r')}{\partial\nu_{r'}}\rho(\vec r')\du \ell'\label{Koper}\\
  S_k\rho(\vec r)=2\int_\Gamma \Phi_k(\vec r,\vec r')\rho(\vec r')\du
  \ell'.
\end{align}
The second term on the right hand side in~\eqref{representationE}
corresponds to the term $\iu\dfrac{k}{2}S_k$ in~\eqref{eq:Eint} and is
added in order to ensure a unique solution for all $k$. The
equation~\eqref{eq:Eint} is often referred to as an indirect combined
field integral equation (ICFIE).

\subsection{H-waves}
We let the magnetic field be parallel to the cylinder, $\vec H(\vec
r)=\hat{\vec z}U(\vec r)$, and let $U(\vec r)=U_\text{inc}(\vec
r)+U_\text{sca}(\vec r)$. The scattered field $U_\text{sca}(\vec r)$
satisfies the following exterior Neumann problem
\begin{align}
&\nabla^2U_\text{sca}(\vec r)+k^2U_\text{sca}(\vec r)=0,\, \vec r\in \Omega_\text{ex}\label{eq3}\\
&\dfrac{\partial U_\text{sca}(\vec r)}{\partial \nu_{\vec r}}=-\dfrac{\partial U_\text{inc}(\vec r)}{\partial \nu_r},\, \vec r\in \Gamma\label{eq4}\\
&\lim_{\vert\vec r\vert\rightarrow \infty}\left(\dfrac{\partial}{\partial r}-\iu k\right)U_\text{sca}(\vec r)=0,
\end{align}
where $\dfrac{\partial U_\text{sca}(\vec r)}{\partial \nu_r}$ is the
normal derivative of $U_\text{sca}$. There are several ways to model
this problem as an integral equation. We use a regularized combined
field integral equation since it is always uniquely solvable. The
scattered field is then obtained from the
representation~\cite{Bruno2012}
\begin{equation}\label{representationH}
U_\text{sca}(\vec r)=\int_\Gamma \Phi(\vec r,\vec r')\rho(\vec r')\du \ell'+\iu\int_\Gamma \dfrac{\partial \Phi(\vec r,\vec r')}{\partial\nu_{r'}}S_{\iu k}\rho(\vec r')\du \ell',\,\vec r\in \Omega_\text{ex},
\end{equation}
which after insertion into~\eqref{eq4} gives the integral equation
\begin{equation}\label{eq67}
(I-K'_{k}-\iu T_kS_{\iu k})\rho(\vec r)=2\dfrac{\partial U_\text{inc}(\vec r)}{\partial \nu_r}.
\end{equation}
Here $K'_k$ is the adjoint to the double layer integral operator $K_k$
in~\eqref{Koper}
\begin{equation}
K'_k\rho(\vec r)=2\int_\Gamma \dfrac{\partial \Phi_k(\vec r,\vec r')}{\partial\nu_{r}}\rho(\vec r')\du \ell'\label{Koperprime}
\end{equation}
and
\begin{equation}\label{Toper}
T_k\rho(\vec r)=\dfrac{\partial}{\partial\nu_{\vec r}}K_k\rho(\vec r).
\end{equation}
The equation~\eqref{eq67} is sometimes referred to as
ICFIE-R~\cite{Bruno2012}.

It is useful to observe that the hypersingular operator $T_k$
in~\eqref{Toper} can be expressed as a sum of a simple operator and an
operator that requires differentiation with respect to arc length
only~\cite{Kress1995}
\begin{equation*}
T_k\rho(\vec r)=2k^2\int_\Gamma \Phi_k(\vec r,\vec r')(\vec{\nu}_{\vec r}\cdot \vec{\nu}_{\vec r'})\rho(\vec r')\du \ell'+2\dfrac{\du}{\du\ell}\int_\Gamma \Phi_k(\vec r,\vec r')\dfrac{\du\rho(\vec r')}{\du\ell'}\du\ell'.
\end{equation*}
We may then rewrite~\eqref{eq67} in the form
\begin{equation}
(I+A_k-\iu B_kS_{\iu k}-\iu C_kC_{\iu k})\rho(\vec r)=2\dfrac{\partial U_\text{inc}(\vec r)}{\partial \nu_r}, \,\,\vec r\in \Gamma,
\label{eq:Hint}
\end{equation}
where $A_k=-K'_k$ and
\begin{align}
&B_k\rho(\vec r)=2k^2\int_\Gamma \Phi_k(\vec r,\vec r')(\vec{\nu}_{\vec r}\cdot \vec{\nu}_{\vec r'})\rho(\vec r)\du\ell'\\
&C_k\rho(\vec r)=2\dfrac{\du}{\du\ell}\int_\Gamma \Phi_k(\vec r,\vec r')\rho(\vec r')\du\ell'.
\end{align}

\section{Numerical scheme}
\label{sec:scheme}

This section briefly reviews the RCIP method, for obtaining accurate
solutions to integral equations on piecewise smooth surfaces, with
focus on basic concepts and on some details particular to the
Helmholtz equation. A richer description, along with demo codes in
{\sc Matlab}, can be found in~\cite{Tutorial2012}.

\subsection{Basics of the RCIP method}

Assume that we have an integral representation of a field $U(\vec r)$,
${\vec r}\in\Omega_{\rm ex}$, in terms of a layer density $\rho(\vec
r)$ on a piecewise smooth boundary $\Gamma$, which leads to a Fredholm
second kind integral equation
\begin{equation}
\left(I+K\right)\rho(\vec r)=g(\vec r)\,,\quad {\vec r}\in \Gamma.
\label{eq:inteq1}
\end{equation}
Here $I$ is the identity, $g$ is a piecewise smooth right hand side,
and $K$ is some integral operator with kernel $K({\vec r},{\vec r}')$
on $\Gamma$ that is compact away from a finite number of corners. Let
us split the kernel
\begin{equation}
K({\vec r},{\vec r}')=
K^{\star}({\vec r},{\vec r}')+K^{\circ}({\vec r},{\vec r}')
\label{eq:ksplit}
\end{equation}
in such a way that $K^{\star}({\vec r},{\vec r}')$ is zero except for
when ${\vec r}$ and ${\vec r}'$ both lie close to the same corner
vertex. In this latter case $K^{\circ}({\vec r},{\vec r}')$ is zero.
The kernel split~(\ref{eq:ksplit}) corresponds to an operator split
\begin{equation}
K=K^{\star}+K^{\circ},
\label{eq:osplit}
\end{equation}
where $K^{\circ}$ is a compact operator. The variable substitution
\begin{equation}
\rho(\vec r)=\left(I+K^{\star}\right)^{-1}\tilde{\rho}(\vec r)
\label{eq:subst}
\end{equation}
allows us to rewrite~(\ref{eq:inteq1}) as a right preconditioned
integral equation
\begin{equation}
\tilde{\rho}(\vec r)+K^{\circ}(I+K^{\star})^{-1}\tilde{\rho}(\vec r)
= g(\vec r)\,,\quad {\vec r}\in \Gamma,
\label{eq:inteq2}
\end{equation}
where the composition $K^{\circ}(I+K^{\star})^{-1}$ is compact.

Let us discretize~(\ref{eq:inteq2}) using a Nystr{\"o}m scheme with
composite 16-point Gauss--Legendre quadrature. The quantities
$\tilde{\rho}$, $K^{\circ}$, and $g$ should be simple to discretize
and resolve accurately on a coarse mesh made of quadrature panels
$\Gamma_p$ of approximately equal length. Only the inverse
$(I+K^{\star})^{-1}$ needs fine local meshes for its accurate
resolution. We arrive at
\begin{equation}
\left({\bf I}_{\rm coa}+{\bf K}_{\rm coa}^{\circ}{\bf R}\right)
\tilde{\boldsymbol{\rho}}_{\rm coa}={\bf g}_{\rm coa},
\label{eq:inteq3}
\end{equation}
where the block-diagonal compressed weighted inverse matrix ${\bf R}$
is given by
\begin{equation}
{\bf R}=
{\bf P}_W^T\left({\bf I}_{\rm fin}+{\bf K}_{\rm fin}^{\star}\right)^{-1}
{\bf P}.
\label{eq:R}
\end{equation}
In~(\ref{eq:inteq3}) and~(\ref{eq:R}) subscript ``coa'' indicates a
grid on the coarse mesh, subscript ``fin'' indicates grids on fine
local meshes, the prolongation matrix ${\bf P}$ performs polynomial
interpolation from the coarse grid to fine grids and ${\bf P}_W^T$ is
the transpose of a weighted prolongation matrix. See~\cite[Section 4
and 5]{Tutorial2012} for details. Once~(\ref{eq:inteq3}) is solved for
$\tilde{\boldsymbol{\rho}}_{\rm coa}$, a discrete weight-corrected
version of the original layer density can be obtained from
\begin{equation}
\hat{\boldsymbol{\rho}}_{\rm coa}={\bf R}
\tilde{\boldsymbol{\rho}}_{\rm coa}.
\end{equation}
The solution $U(\vec r)$ can then be recovered in most of the
computational domain using $\hat{\boldsymbol{\rho}}_{\rm coa}$ in a
discretized version of the integral representation for $U(\vec r)$.

%\eqref{eq:inteq3}
%~(\ref{eq:inteq3}), 
Note that in \eqref{eq:inteq3}, the need for resolution in corners is
not visible. The transformed layer density
$\tilde{\boldsymbol{\rho}}_{\rm coa}$ on a non-smooth $\Gamma$ should
be as easy to solve for as the original layer density
$\boldsymbol{\rho}_{\rm coa}$ in a discretization of~(\ref{eq:inteq1})
on a smooth $\Gamma$. All computational difficulties are concentrated
in the matrix ${\bf R}$. Let there be $n$ discretization points on the
local fine grid close to a particular corner on $\Gamma$. Judging from
the definition~(\ref{eq:R}), it seems as if computing ${\bf R}$ should
be a prohibitively expensive and also unstable undertaking for large
$n$. Fortunately, ${\bf R}$ can be computed via a fast and stable
recursion which relies on a hierarchy of small nested meshes.  This
fast recursion enables the computation of the diagonal block of ${\bf
  R}$, that corresponds to a particular corner, at a cost only
proportional to $n$. Actually, when very large $n$ are needed for
resolution the cost can be further cut down with the use of Newton's
method. See~\cite[Section 6 and 12]{Tutorial2012} for details.

The fast recursion for ${\bf R}$ can also be run backwards for the
purpose of reconstructing $\boldsymbol{\rho}_{\rm fin}$ from
$\tilde{\boldsymbol{\rho}}_{\rm coa}$. A partial reconstruction of
$\boldsymbol{\rho}_{\rm fin}$ is needed when $U(\vec r)$ is to be
evaluated at points in $\Omega_{\rm ex}$ that lie close to corner
vertices. See~\cite[Section 9]{Tutorial2012} for details.

We remark that the integral equations~\eqref{eq:Eint}
and~\eqref{eq:Hint}, which are to be solved in this paper, have a more
complicated appearance than the model equation~(\ref{eq:inteq1}). In
practice this poses no problems for RCIP -- just some extra work. The
two integral operators in~\eqref{eq:Eint} can, for programming
purposes, be combined into a single operator. The composition of
integral operators in~\eqref{eq:Hint} can be treated with an expansion
technique. With the help of two new temporary layer densities, one can
arrive at a recursion for an expanded compressed inverse matrix ${\bf
  R}$ with the same structure as~(\ref{eq:R}). Once ${\bf R}$ is
computed one can extract separate blocks from it and use them in a
more involved version of~(\ref{eq:inteq3}) that still uses only a
single transformed global density $\tilde{\boldsymbol{\rho}}_{\rm
  coa}$. See~\cite[Section 14 and 17]{Tutorial2012} for details.

\subsection{The discretization of Hankel kernels}\label{sec:discretization}

High-order accurate Nystr{\"o}m discretization of boundary integral
equations associated with the Helmholtz equation is a topic that has
received much attention recently. See~\cite{Hao2011} for a comparison
of various 2D schemes. We now present our preferred scheme by showing
how to discretize the operator $K_k$ of~(\ref{Koper}) and the first
operator on the right hand side of~(\ref{representationE}). The other
integral operators of Section~\ref{sec:form} are discretized in
similar ways.

The kernel of $K_k$ is twice that of the first operator
in~(\ref{representationE}) and can, modulo a constant of ${\rm i}/2$,
be expressed as
\begin{equation}
K_k(\vec r,\vec r')=k|\vec r-\vec r'| H^{(1)}_1 (k|\vec r-\vec r'|)
\frac{(\vec r -\vec r')\cdot\nu_{r'}}{|\vec r-\vec r'|^2},
\label{eq:Kkkern}
\end{equation}
where $H^{(1)}_1$ is the Hankel function of the first kind of order
one. When $\vec r\in\Gamma$, it is instructive to
write~(\ref{eq:Kkkern}) in the form
\begin{equation}
K_k(\vec r,\vec r')=f(\vec r,\vec r')+\frac{2{\rm i}}{\pi}
\log|\vec r-\vec r'|\Re\left\{K_k(\vec r,\vec r')\right\}.
\label{eq:twosmooth}
\end{equation}
For a fixed $\vec r\in\Gamma$, we see from~(\ref{eq:Kkkern}) and a
series representation of $H^{(1)}_1$ that $f(\vec r,\vec r')$ and
$\Re\left\{K_k(\vec r,\vec r')\right\}$ are smooth functions of $\vec
r'\in\Gamma$ and that
\begin{equation}
\lim_{\vec r'\to\vec r}
\log|\vec r-\vec r'|\Re\left\{K_k(\vec r,\vec r')\right\}=0.
\end{equation}

Consider now the integral $I_p(\vec r)$ over a quadrature panel
$\Gamma_p$
\begin{equation}
I_p(\vec r)=\int_{\Gamma_p}K_k(\vec r,\vec r')\rho(\vec r')\,\du \ell'.
\label{eq:Ip}
\end{equation}
Let $\vec r(t)$ be a parameterization of $\Gamma$. Discretizing $K_k$
means being able to evaluate~(\ref{eq:Ip}) for all $\vec r$ of
interest, given a set of values $\rho(\vec r(t_j))$ on each
$\Gamma_p$.

If $\vec r$ is a point away from $\Gamma_p$, then $K_k(\vec r,\vec
r')$ is a smooth function of $\vec r'\in\Gamma_p$ and $I_p(\vec r)$
can be evaluated to high accuracy using 16-point Gauss--Legendre
quadrature
\begin{equation}
I_p(\vec r)\approx\sum_jK_k(\vec r,\vec r_j)\rho_js_jw_j,
\label{eq:disc1}
\end{equation}
where $\vec r_j=\vec r(t_j)$, $\rho_j=\rho(\vec r(t_j))$, $s_j=|{\rm
  d}\vec r(t_j)/{\rm d}t|$, and $t_j$ and $w_j$ are nodes and weights
on $\Gamma_p$.

If $\vec r_i$ is a discretization point close to $\Gamma_p$ or on
$\Gamma_p$, then $K_k(\vec r_i,\vec r')$ is not a (sufficiently)
smooth function of $\vec r'\in\Gamma_p$ and we
use~(\ref{eq:twosmooth}) to arrive at
\begin{equation}
I_p(\vec r_i)\approx
\sum_jf(\vec r_i,\vec r_j)\rho_js_jw_j
+\frac{2{\rm i}}{\pi}\sum_j
\Re\left\{K_k(\vec r_i,\vec r_j)\right\}\rho_jw_{ij{\rm L}},
\label{eq:log1}
\end{equation}
where $w_{ij{\rm L}}$ are high-order product integration weights for
the logarithmic operator which can be constructed using the analytic
method in~\cite[Section 2.3]{HelsingJCP2009}. The
formula~(\ref{eq:log1}) can be rearranged into a particularly
convenient form
\begin{equation}
I_p(\vec r_i)\approx
\sum_jK_k(\vec r_i,\vec r_j)\rho_js_jw_j
+\frac{2{\rm i}}{\pi}\sum_j
\Re\left\{K_k(\vec r_i,\vec r_j)\right\}\rho_js_jw_j
w_{ij{\rm L}}^{\rm corr},
\label{eq:disc2}
\end{equation}
where the weight corrections
\begin{equation}
w_{ij{\rm L}}^{\rm corr}=
\left(\frac{w_{ij{\rm L}}}{s_jw_j}-\log|\vec r_i-\vec r_j|\right)
\label{eq:logcorr}
\end{equation}
are cheap to compute and depend only on the relative length (in
parameter) of neighboring quadrature panels and on nodes and weights
on a canonical panel. The formula~(\ref{eq:disc1}) with $\vec r=\vec
r_i$ and~(\ref{eq:disc2}) summarize our Nystr{\"o}m discretization of
$K_k$ on $\Gamma$.

If $\vec r$ is a point not on $\Gamma$ but in $\Omega_{\rm ex}$ close
to $\Gamma_p$, we write~(\ref{eq:Kkkern}) in the form
\begin{equation}
K_k(\vec r,\vec r')=g(\vec r,\vec r')
+\frac{2{\rm i}}{\pi}\log|\vec r-\vec r'|\Re\left\{K_k(\vec r,\vec r')\right\}
+\frac{2{\rm i}}{\pi}\frac{(\vec r'-\vec r)\cdot\nu_{r'}}{|\vec r'-\vec r|^2}.
\label{eq:threesmooth}
\end{equation}
We see from~(\ref{eq:Kkkern}) and a series representation of
$H^{(1)}_1$ that $g(\vec r,\vec r')$ and $\Re\left\{K_k(\vec r,\vec
  r')\right\}$ are smooth functions of $\vec r'$. In analogy
with~(\ref{eq:disc2}) one can write
\begin{multline}
I_p(\vec r)\approx
\sum_jK_k(\vec r,\vec r_j)\rho(\vec r_j)s_jw_j
+\frac{2{\rm i}}{\pi}\sum_j
\Re\left\{K_k(\vec r,\vec r_j)\right\}\rho(\vec r_j)s_jw_j
w_{j{\rm L}}^{\rm corr}(\vec r)\\
+\frac{2{\rm i}}{\pi}\sum_j\rho(\vec r_j)
\left(w_{j{\rm C}}(\vec r)-
\frac{(\vec r_j-\vec r)\cdot\nu_{r_j}}{|\vec r_j-\vec r|^2}s_jw_j\right),
\label{eq:disc3}
\end{multline}
where $w_{j{\rm L}}^{\rm corr}(\vec r)$ are weight corrections as
in~(\ref{eq:logcorr}), but with $\vec r_i$ replaced by $\vec r$, and
$w_{j{\rm C}}(\vec r)$ are high-order product integration weights for
the Cauchy singular operator which can be constructed using the
analytic method in~\cite[Section 2.1]{HelsingJCP2009}. The
formulas~(\ref{eq:disc1}) and~(\ref{eq:disc3}) are used to discretize
the first operator in~(\ref{representationE}) when producing field
plots.

\subsection{Convergence and error estimates}

Our solver shows a stable behavior. This means that the solution
converges rapidly with coarse mesh refinement up until a point beyond
which no further improvement occurs. Actually, beyond this optimal
point there will be a slow decay in the quality of the solution, due
to accumulated roundoff error. The precise location of the optimal
point is hard to determine {\it a priori}. It depends on the geometry,
on the boundary conditions, and on the wave number. The optimal point
is determined experimentally in the numerical examples of
Section~\ref{sec:examples}.

We have estimated the accuracy in our solutions $U(\vec r)$ rather
thoroughly. The tutorial~\cite[Section 18]{Tutorial2012} contains
error plots for exterior problems in non-smooth domains produced in a
direct way. These are achieved by generating the boundary conditions
on $\Gamma$ via line sources inside $\Gamma$ so that the exact
solution is known. In the plane-wave scattering examples of
Section~\ref{sec:near}, below, no exact results are known. Therefore
we proceed as follows: we first compute a solution $U(\vec r)$ using a
number of coarse panels on $\Gamma$ deemed sufficient for resolution.
Then we increase this number with 50 \% and solve again. The
difference between the resolved value of $U(\vec r)$ and the
overresolved value of $U(\vec r)$ is used as an indirect pointwise
error estimate. Yet an indirect method to estimate the (overall)
precision in the computations is by comparing the scattering cross
section computed from its definition (close to $\Gamma$) with its
value obtained via the optical theorem (at infinity). See, further,
Section~\ref{sec:cross}. As it turns out, the various error estimates
seem to agree well.

\section{Numerical examples}
\label{sec:examples}

\begin{figure}
\centering
\includegraphics[height=55mm]{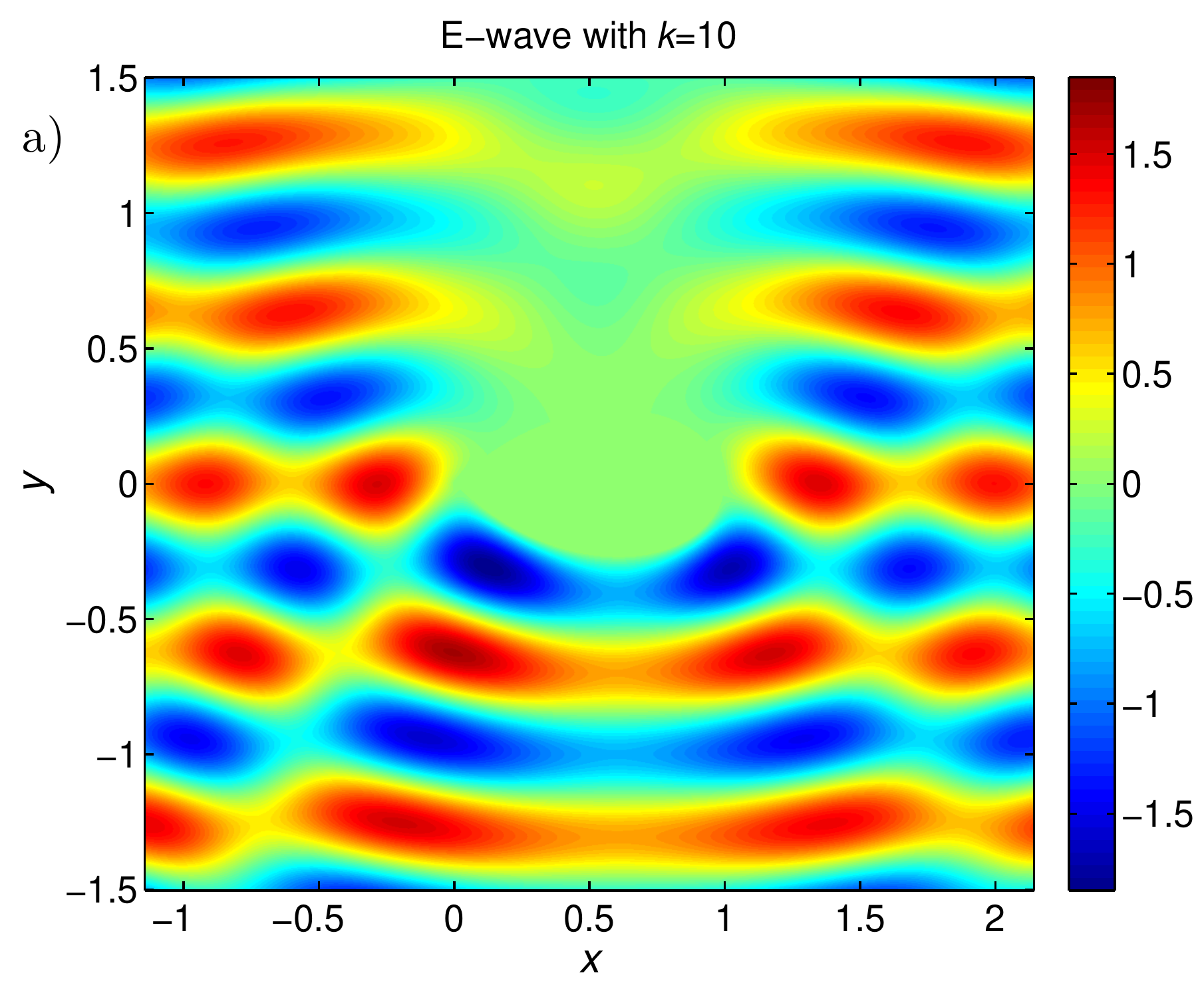}
\includegraphics[height=55mm]{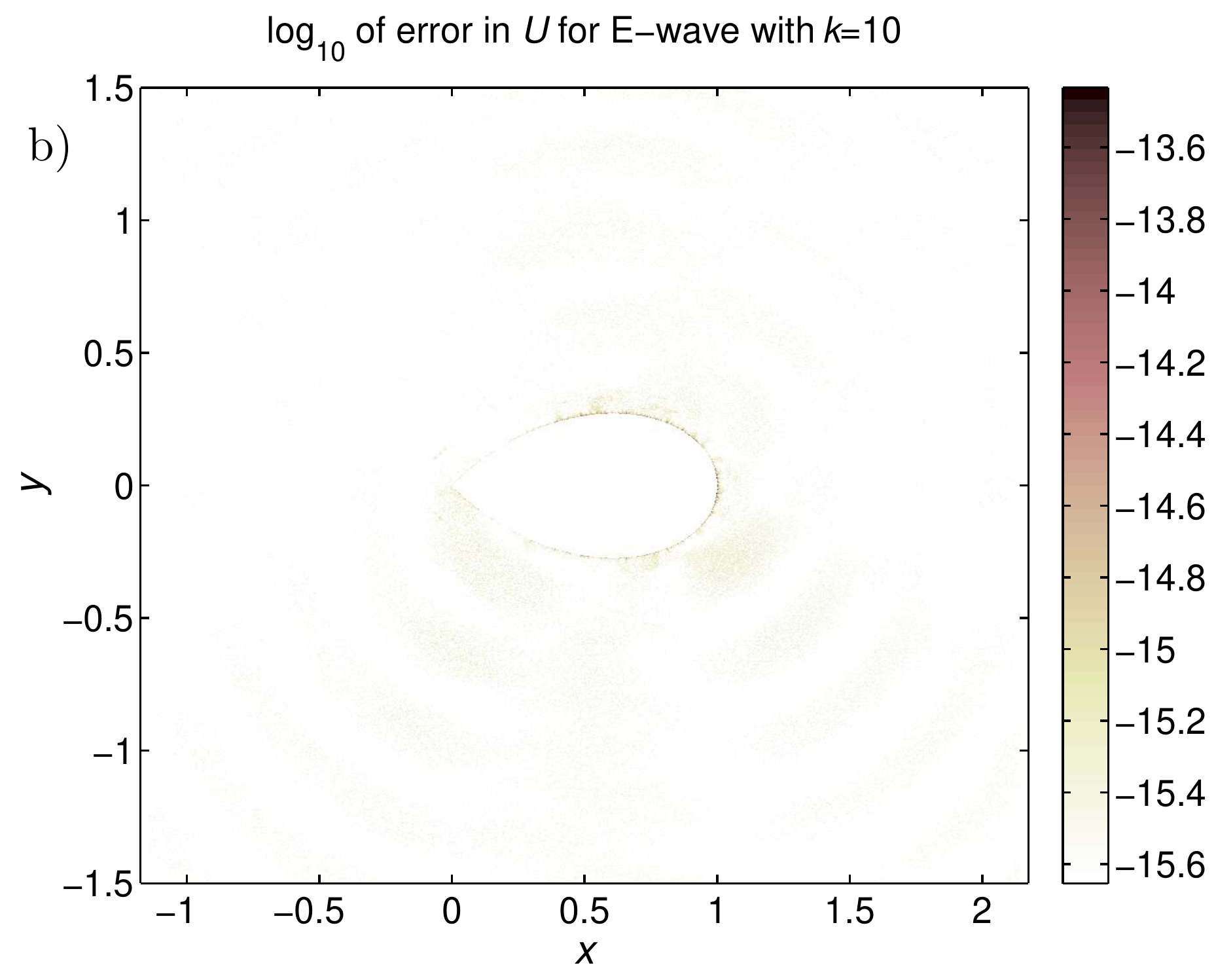}
\includegraphics[height=54mm]{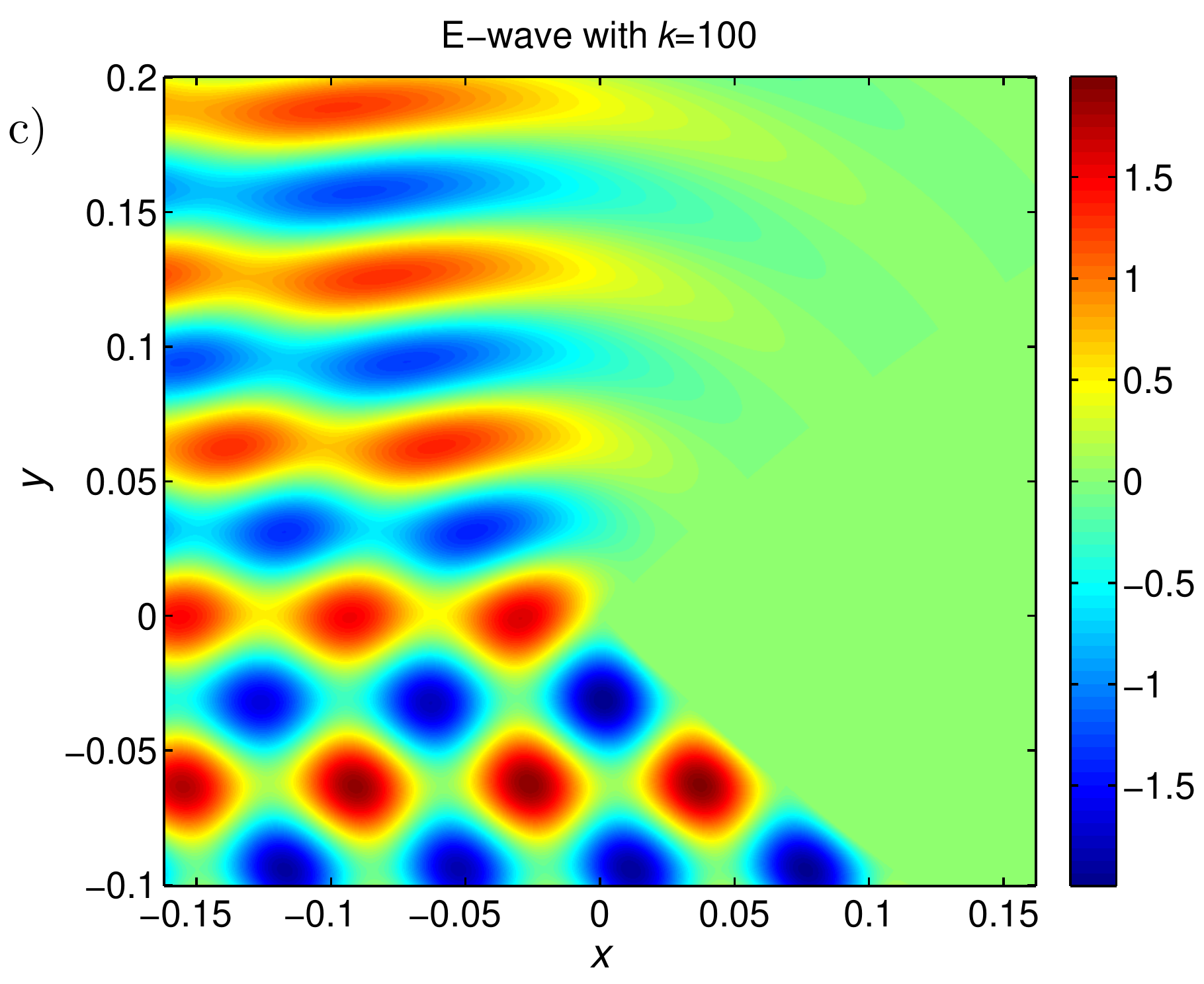}
\includegraphics[height=54mm]{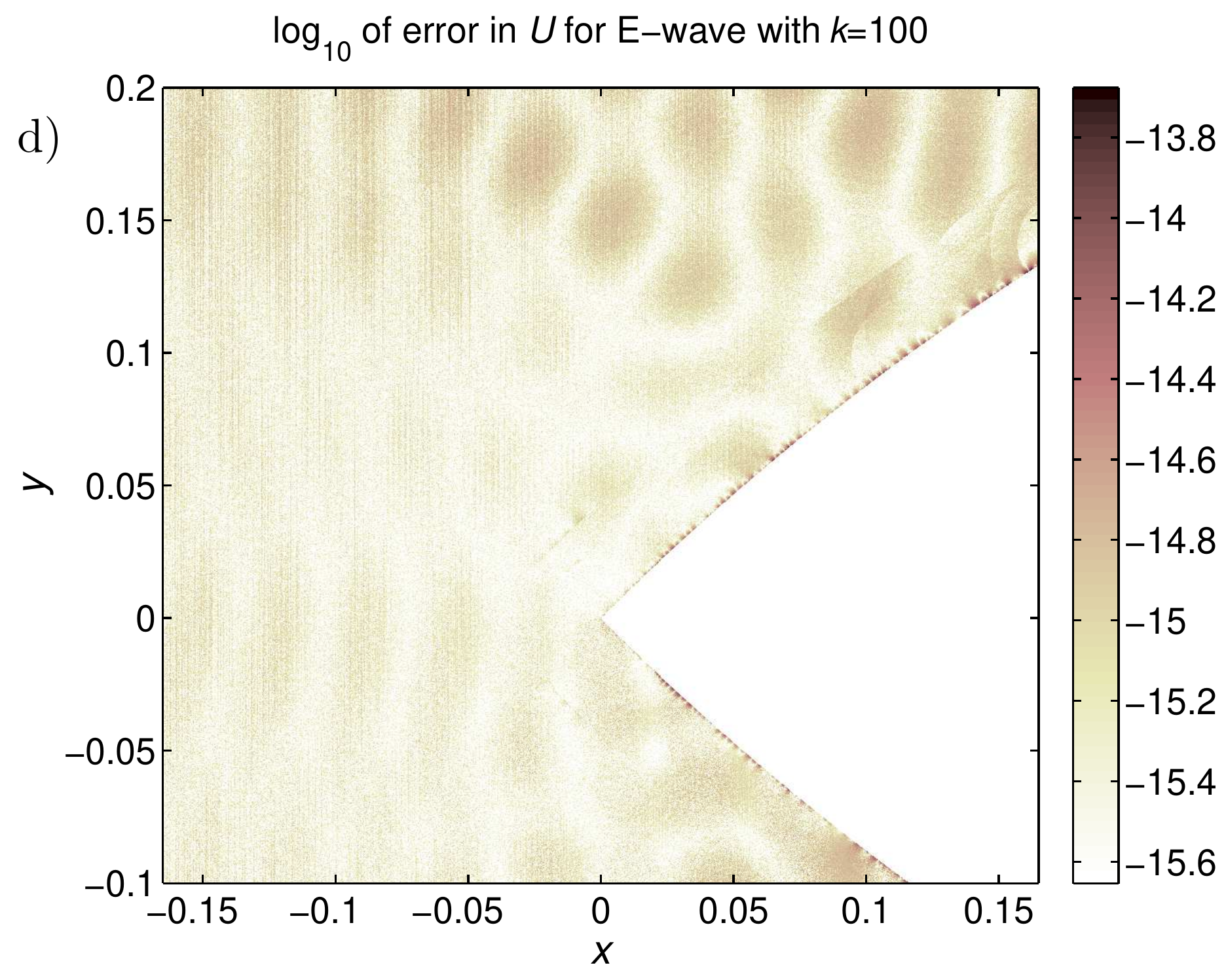}
\includegraphics[height=55mm]{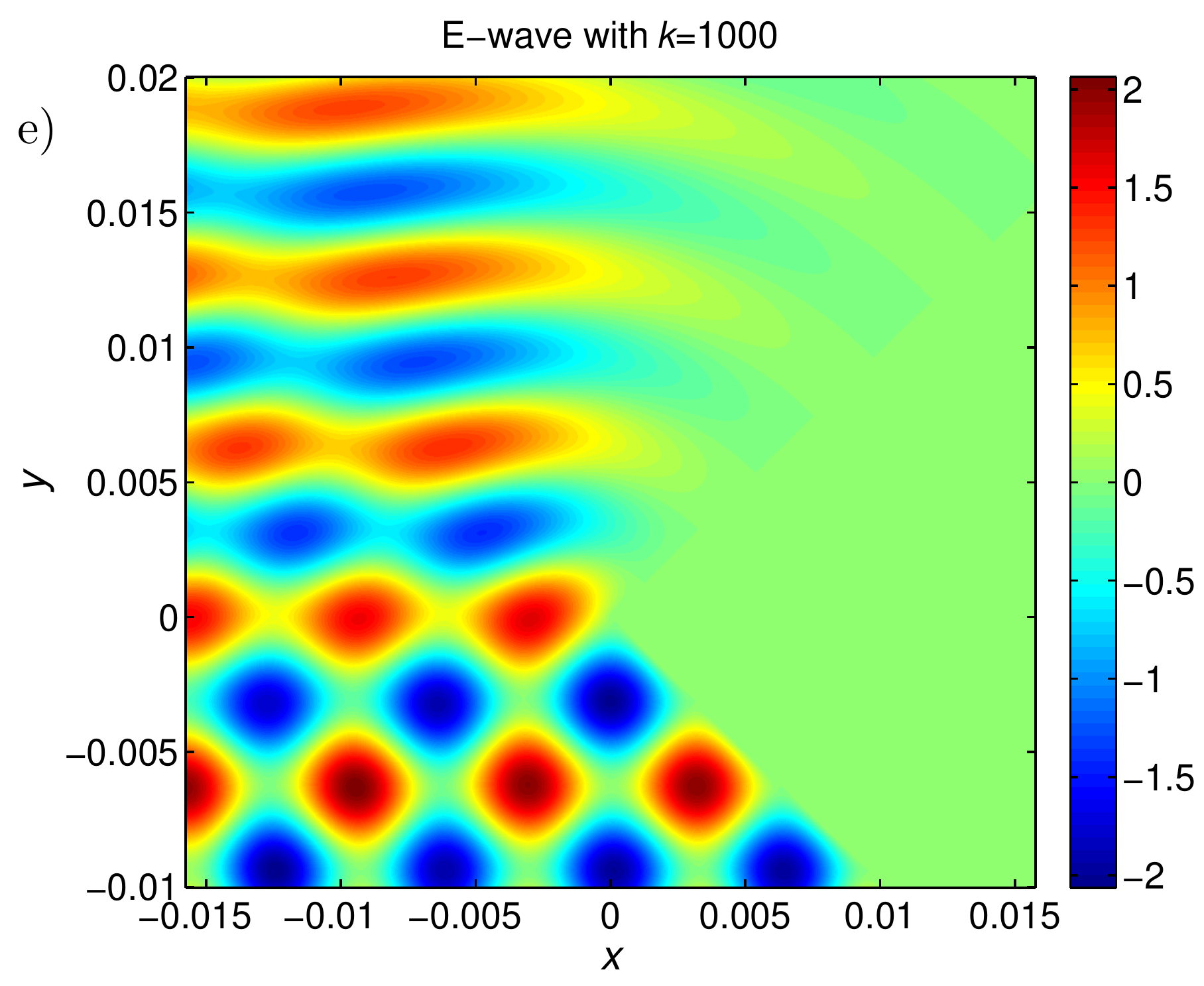}
\includegraphics[height=55mm]{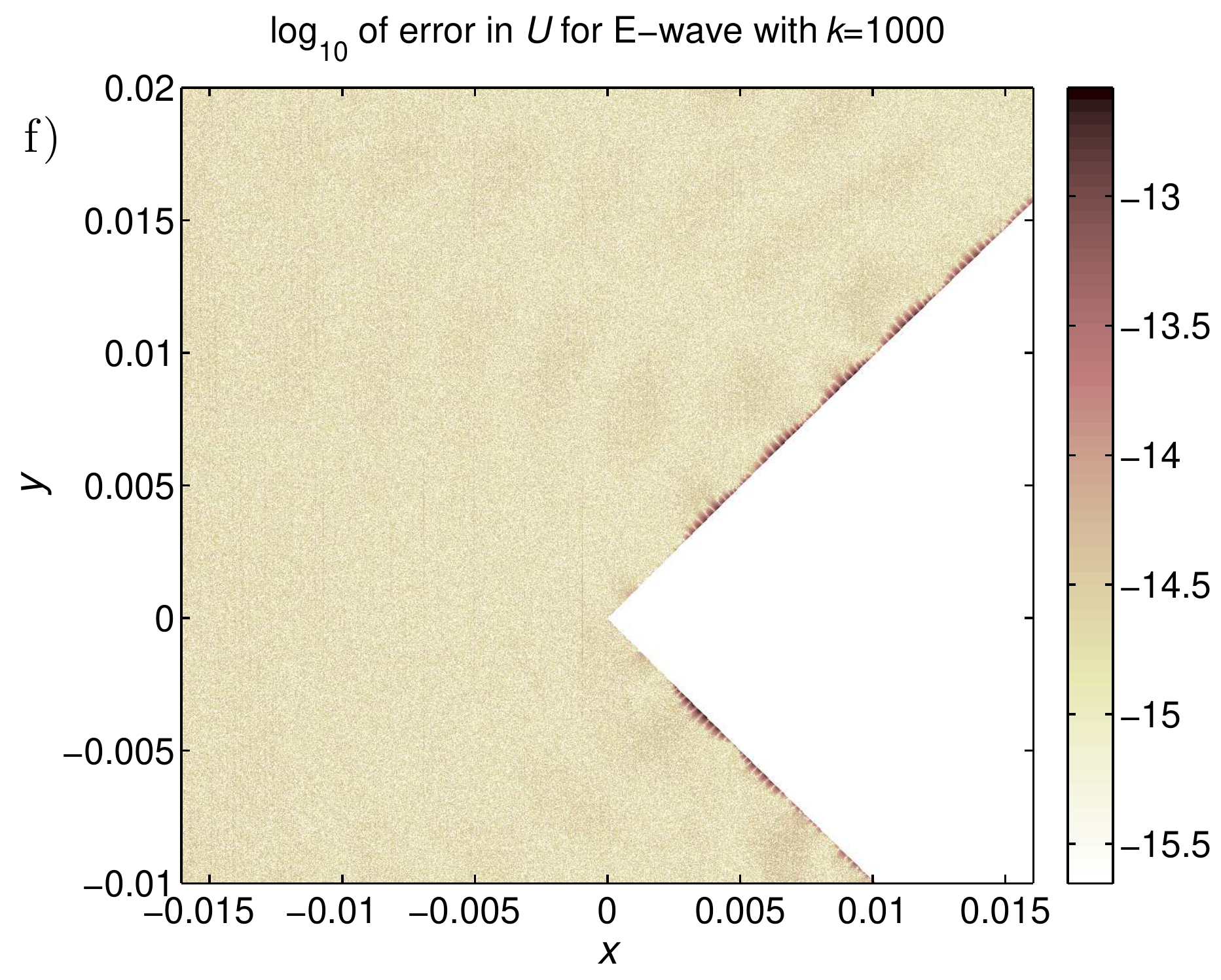}
\caption{Left: a), c), e) show $\Re\left\{U(\vec r)\right\}$
  for a plane E-wave $U_\text{inc}(\vec r)=e^{\iu k y}$  incident on the perfectly
  conducting cylinder with boundary $\Gamma$ given by~\eqref{eq:gamma}.
  Right: b), d), f) show absolute errors.}
\label{e-wave}
\end{figure}

\begin{figure}
\centering
\includegraphics[height=55mm]{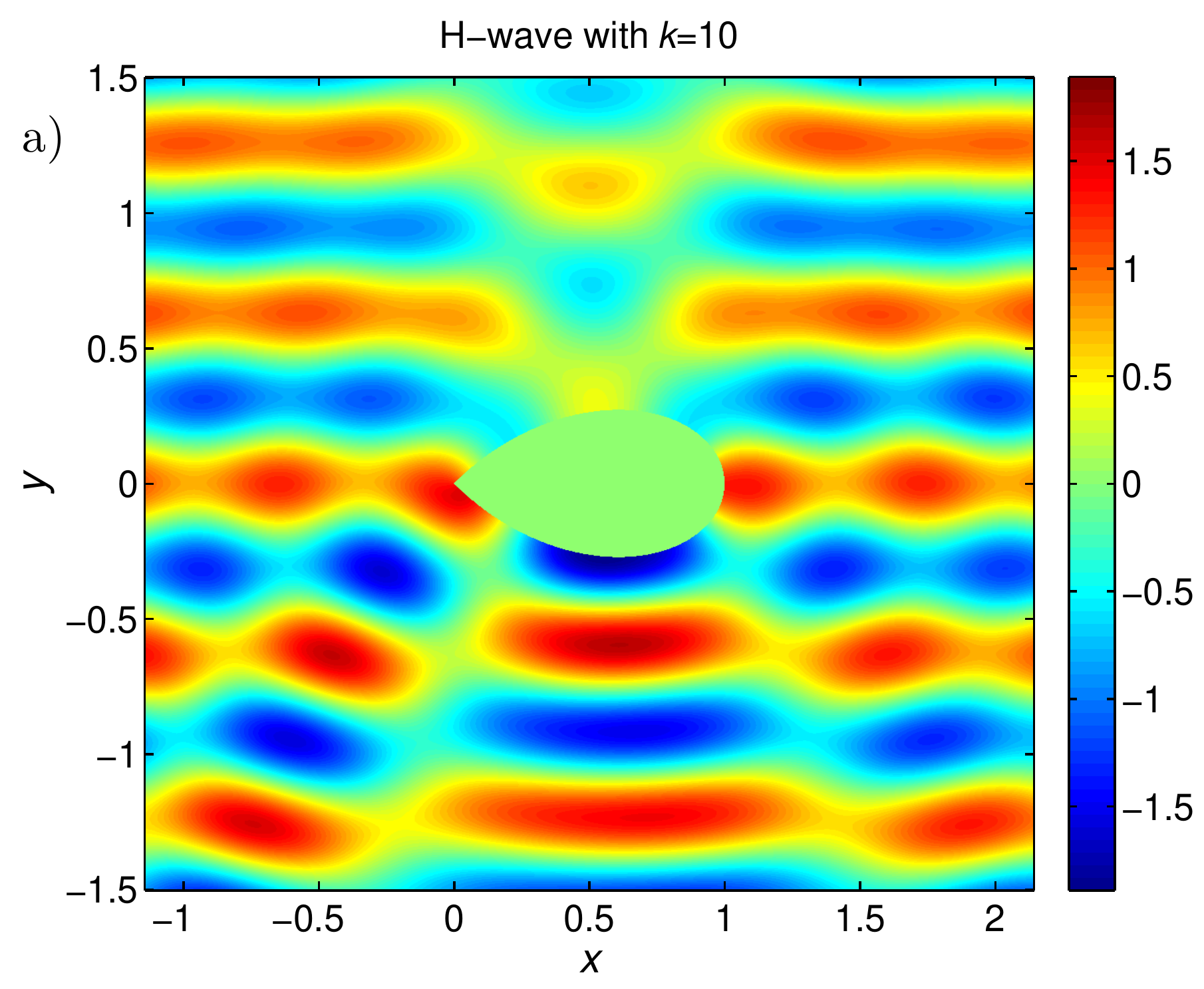}
\includegraphics[height=55mm]{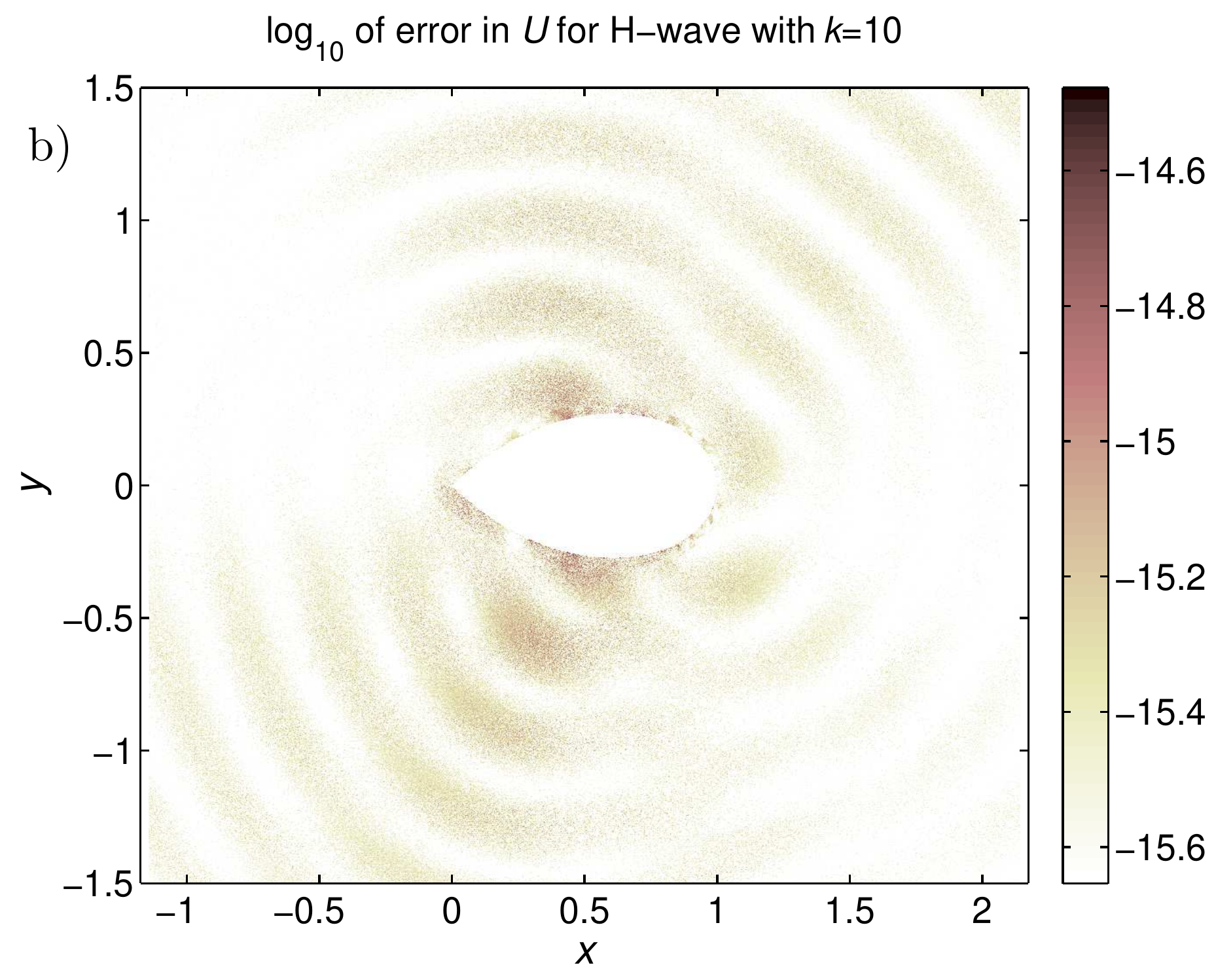}
\includegraphics[height=54mm]{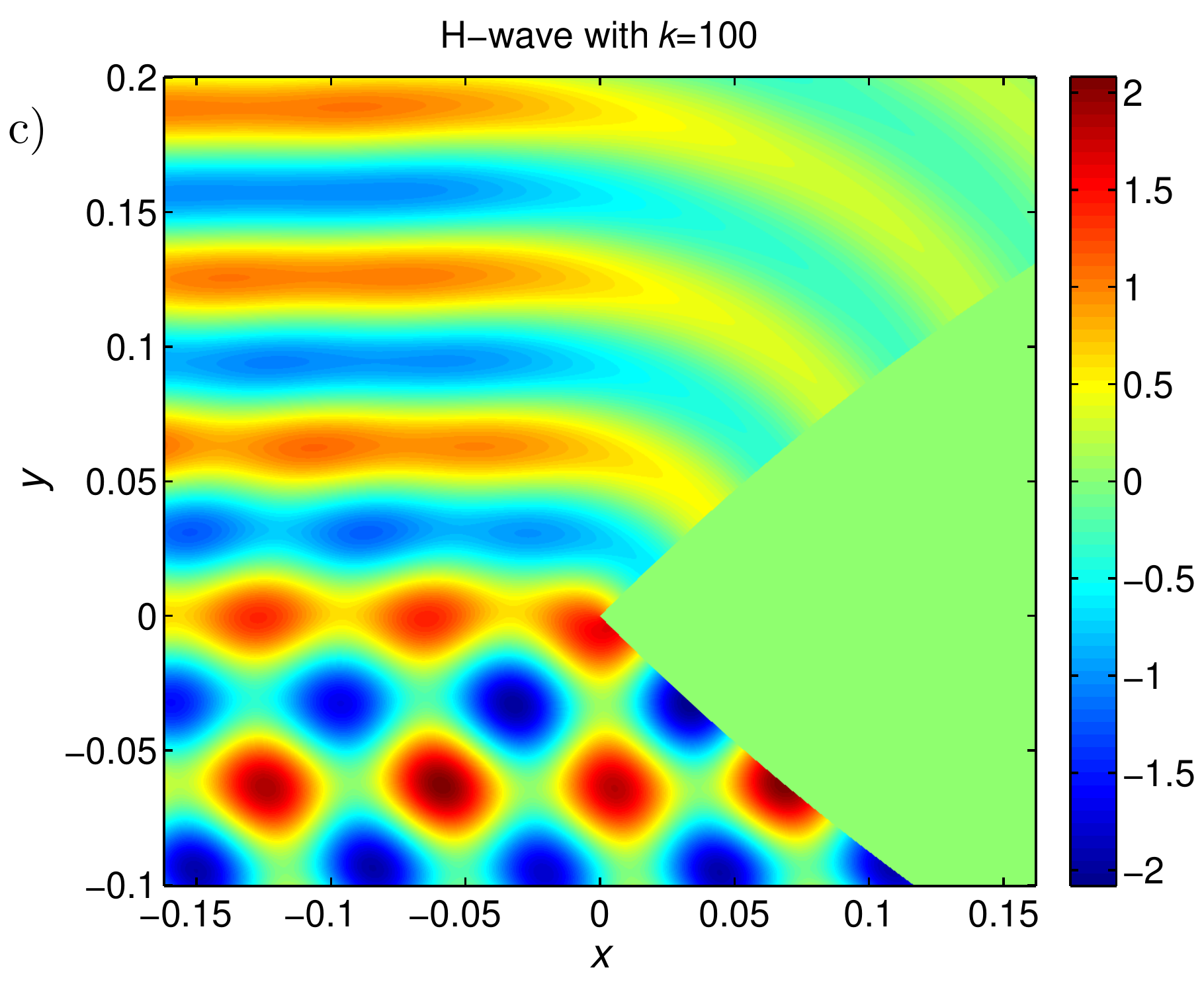}
\includegraphics[height=54mm]{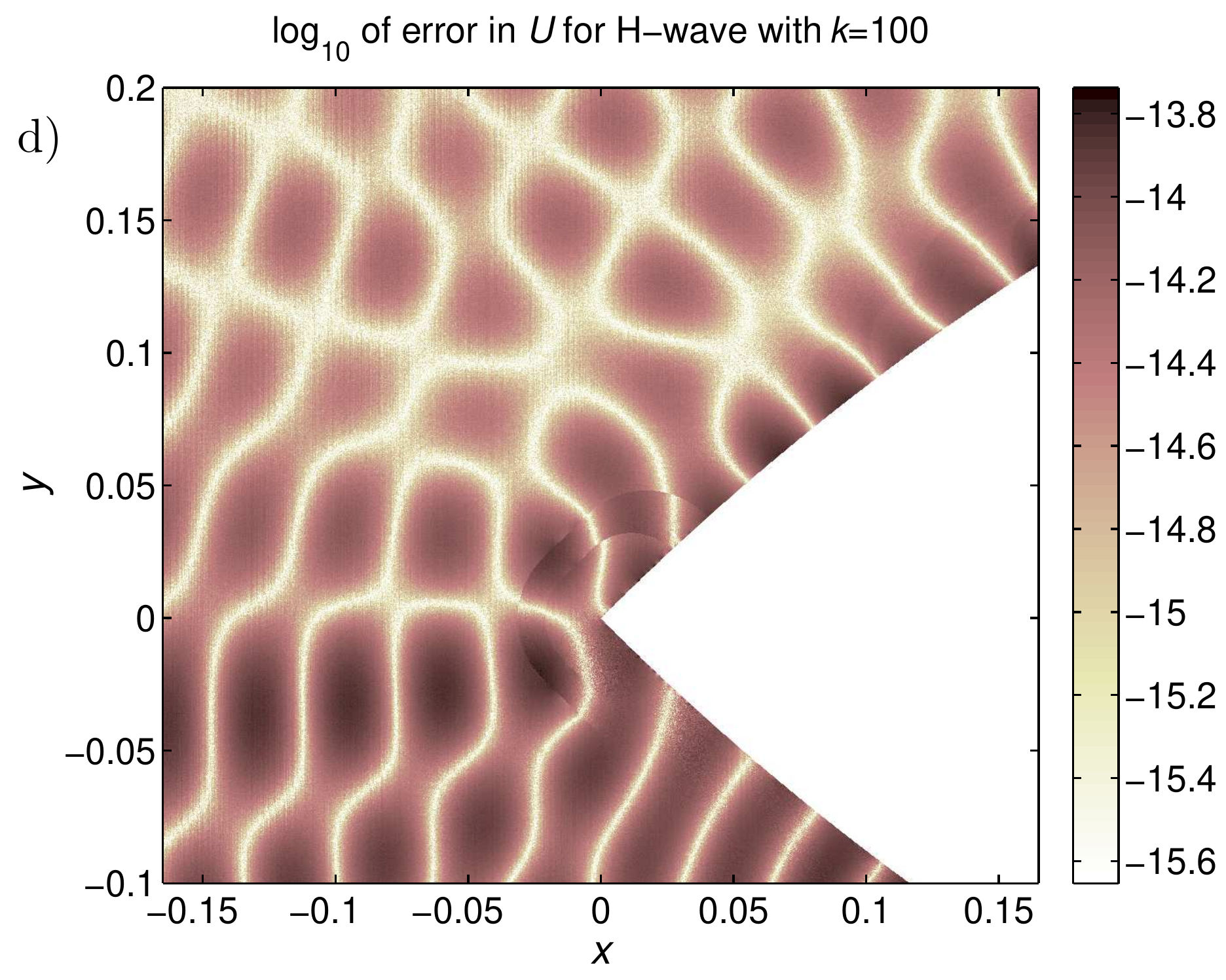}
\includegraphics[height=55mm]{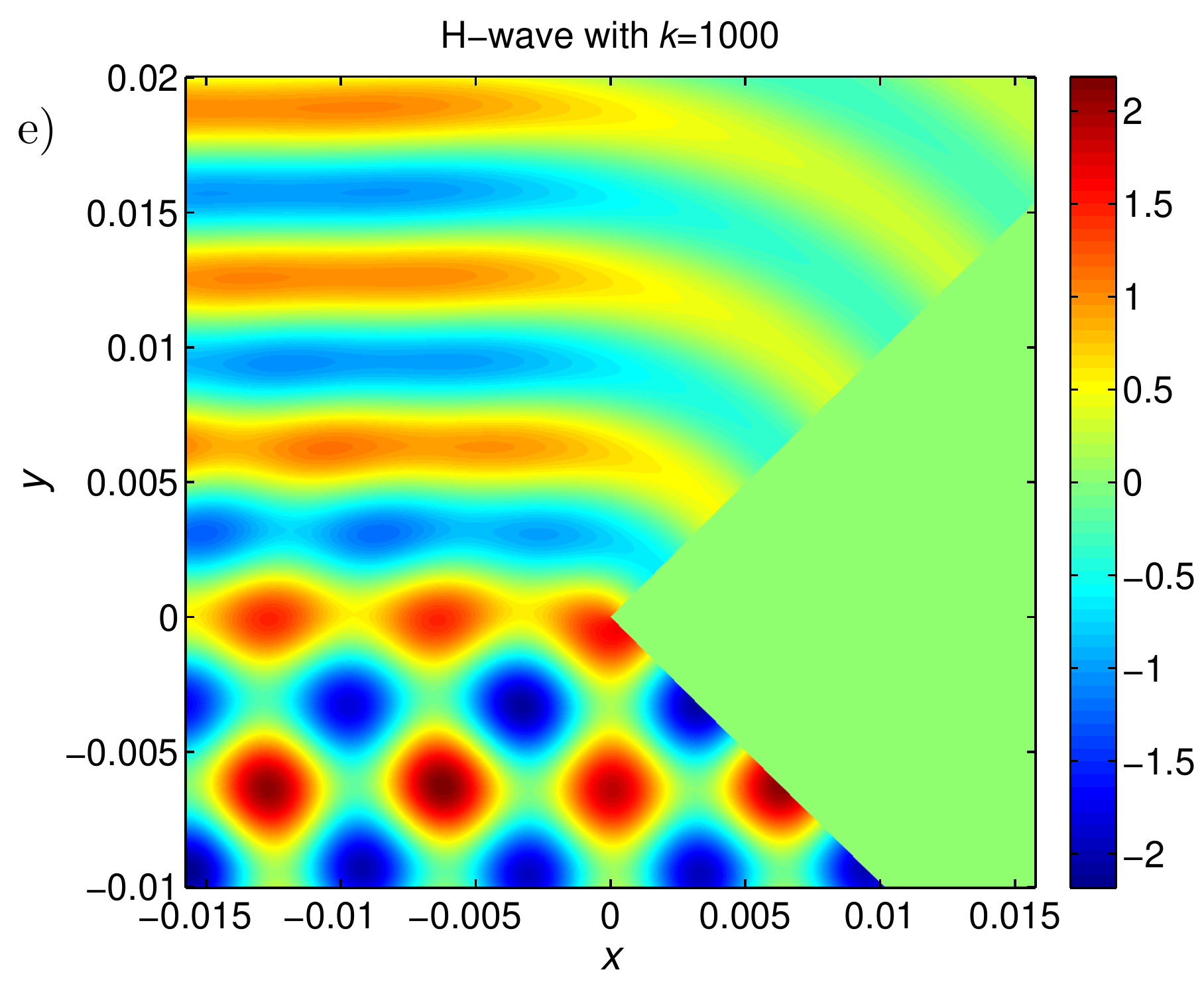}
\includegraphics[height=55mm]{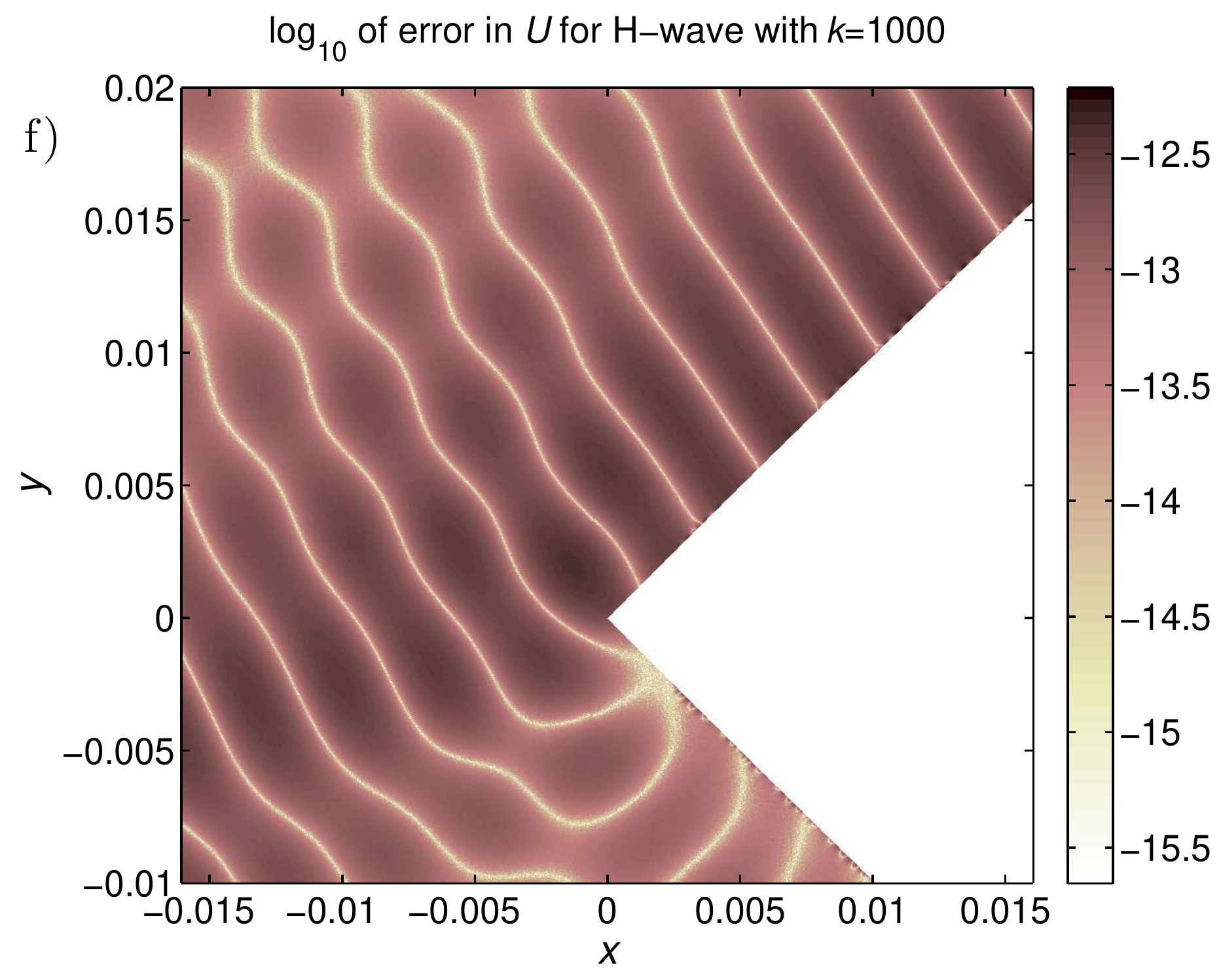}
\caption{Left: a), c), e) show $\Re\left\{U(\vec r)\right\}$
  for a plane H-wave $U_\text{inc}(\vec r)=e^{\iu k y}$  incident on the perfectly
  conducting cylinder with boundary $\Gamma$ given by~\eqref{eq:gamma}.
  Right: b), d), f) show absolute errors.}
\label{h2-wave}
\end{figure}

We shall now solve~\eqref{eq:Eint} and~\eqref{eq:Hint} for the unknown
density $\rho(\vec r)$, using the method of Section~\ref{sec:scheme},
and then evaluate the scattered fields of~\eqref{representationE}
and~\eqref{representationH}. We restrict the numerical examples to
scattering from an infinite straight cylinder with boundary $\Gamma$
described by
\begin{equation}
{\vec r}(t)=\sin(\pi t)\left(\cos((t-0.5)\pi/2),\sin((t-0.5)\pi/2)\right)\,,
\quad t\in[0,1]\,,
\label{eq:gamma}
\end{equation}
and to the incident plane wave $U_\text{inc}(\vec r)=e^{\iu k y}$ for
both E-waves and H-waves. The object parameterized in \eqref{eq:gamma}
has a corner with opening angle $\theta=\pi/2$ at ${\vec r}=0$ and a
diameter $d=1$, in arbitrary length units, so that $kd=k$. The
examples cover sizes from $kd=1$ up to $kd=1000$. We have seen that at
$kd=1000$ the frequency is high enough such that the uniform theory of
diffraction theory can be applied. All numerical examples are executed
in {\sc MATLAB} on a workstation equipped with an IntelXeon E5430 CPU
at 2.66 GHz and 32 GB of memory.

\subsection{Near field}
\label{sec:near}

A criterion for a powerful method is that it should be able to
calculate the electric and magnetic fields everywhere in
$\Omega_\text{ex}$. Figures \ref{e-wave} and \ref{h2-wave} show the
total electric field for the E-wave and total magnetic field for the
H-wave in the vicinity of the scattering object and the corresponding
errors. The scattering object itself appears in green color in the
left images and in white color in the right images. The number of
spatial points in each image is $10^6$. It is encouraging to see, in
the right images of Figures \ref{e-wave} and \ref{h2-wave}, that the
accuracy is high even close to the boundary and, in particular, close
to the corner. The integrals in \eqref{representationE} and
\eqref{representationH} are often thought of as difficult to evaluate
close to the boundary due to the singularities in the Hankel functions
when $\vec r'=\vec r$. However, the present method circumvents these
problems using the high-order analytic quadrature outlined in Section
\ref{sec:discretization}.

In Figures \ref{e-wave} a), c), e) the real part of the total electric
field $U(\vec r)$ for the E-wave case is plotted for $kd=10,\, 100$,
and 1000. To capture the diffraction pattern in the vicinity of the
corner, the field is plotted in a rectangular region with side length
proportional to $1/k$ and center at the tip of the corner. At $kd=10$
the error is very small, as seen from Figure \ref{e-wave} b). The
errors increase slightly with $kd$ but even at $kd=1000$ we get 14
digits or better almost everywhere, as depicted in Figure \ref{e-wave}
f).  For H-waves the accuracy is almost as good as for the E-waves, as
seen from Figure \ref{h2-wave}.

For $kd=100$ and 1000 we can interpret the field plots in
Figures~\ref{e-wave} c), e) and ~\ref{h2-wave} c), e) through the
theory of diffraction. Thus, the outer region $\Omega_\text{ex}$ is
divided into three subregions separated by the reflection boundary and
the shadow boundary.

\begin{figure}
\centering
\includegraphics[height=55mm]{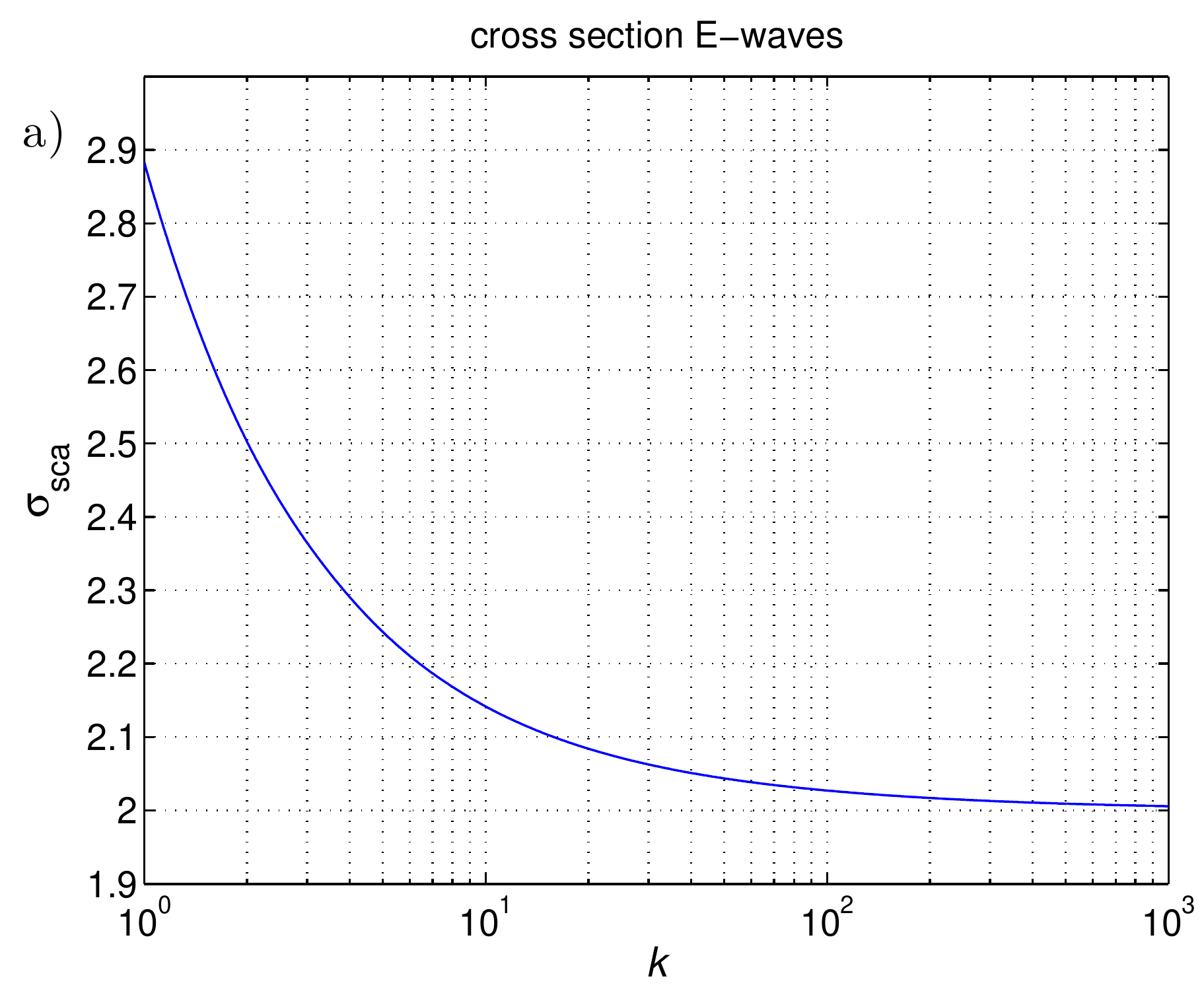}
\includegraphics[height=55mm]{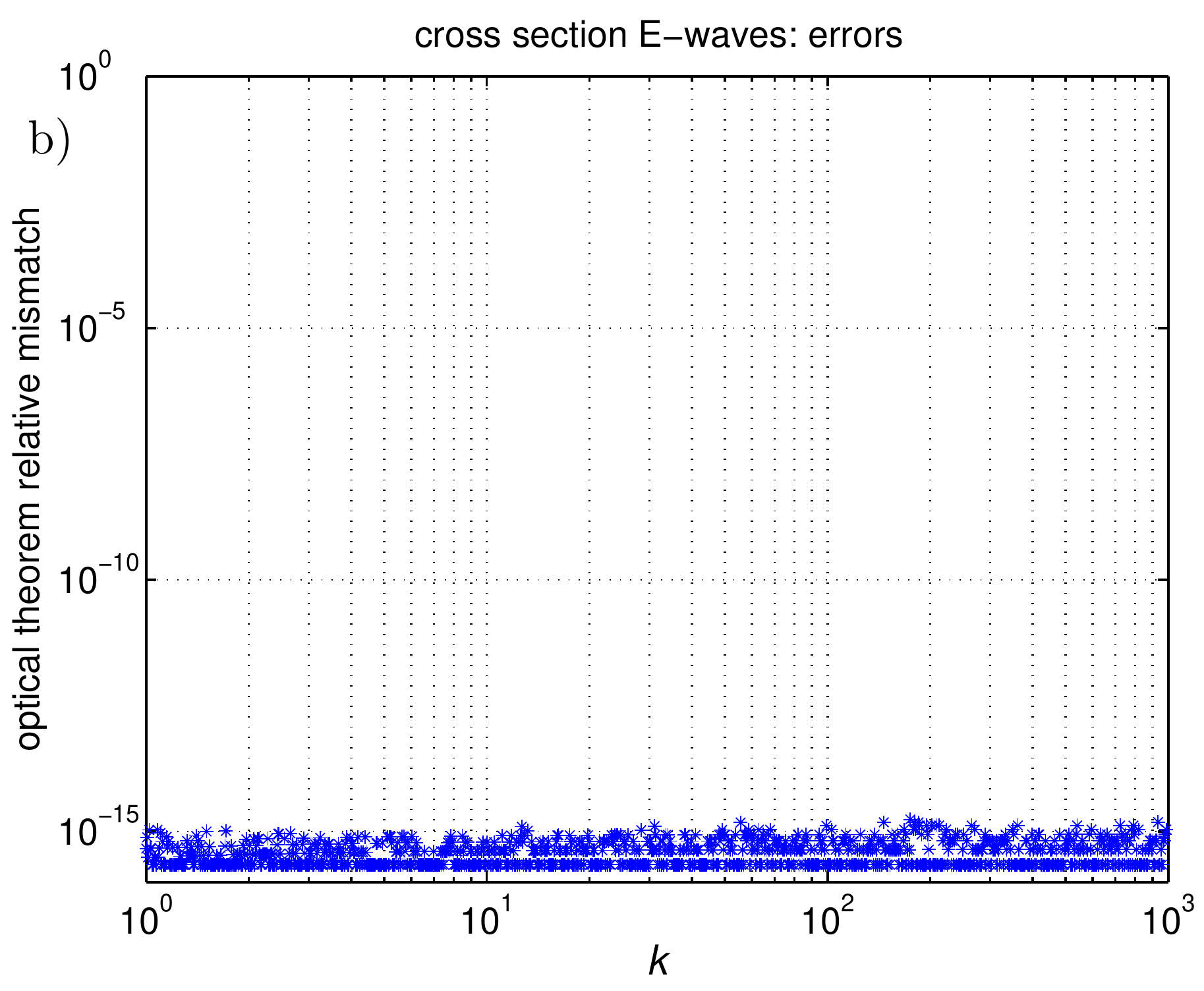}
\includegraphics[height=55mm]{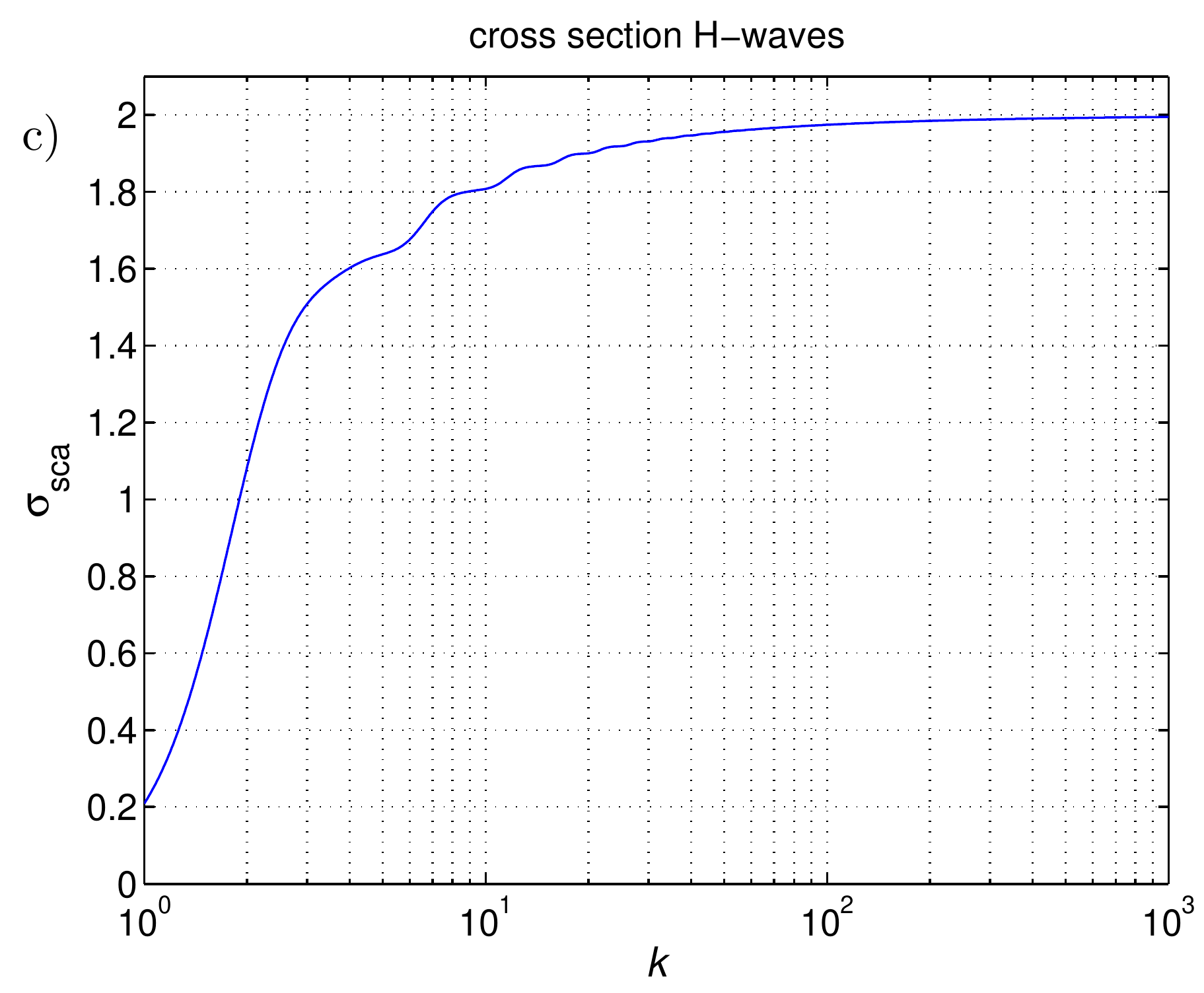}
\includegraphics[height=55mm]{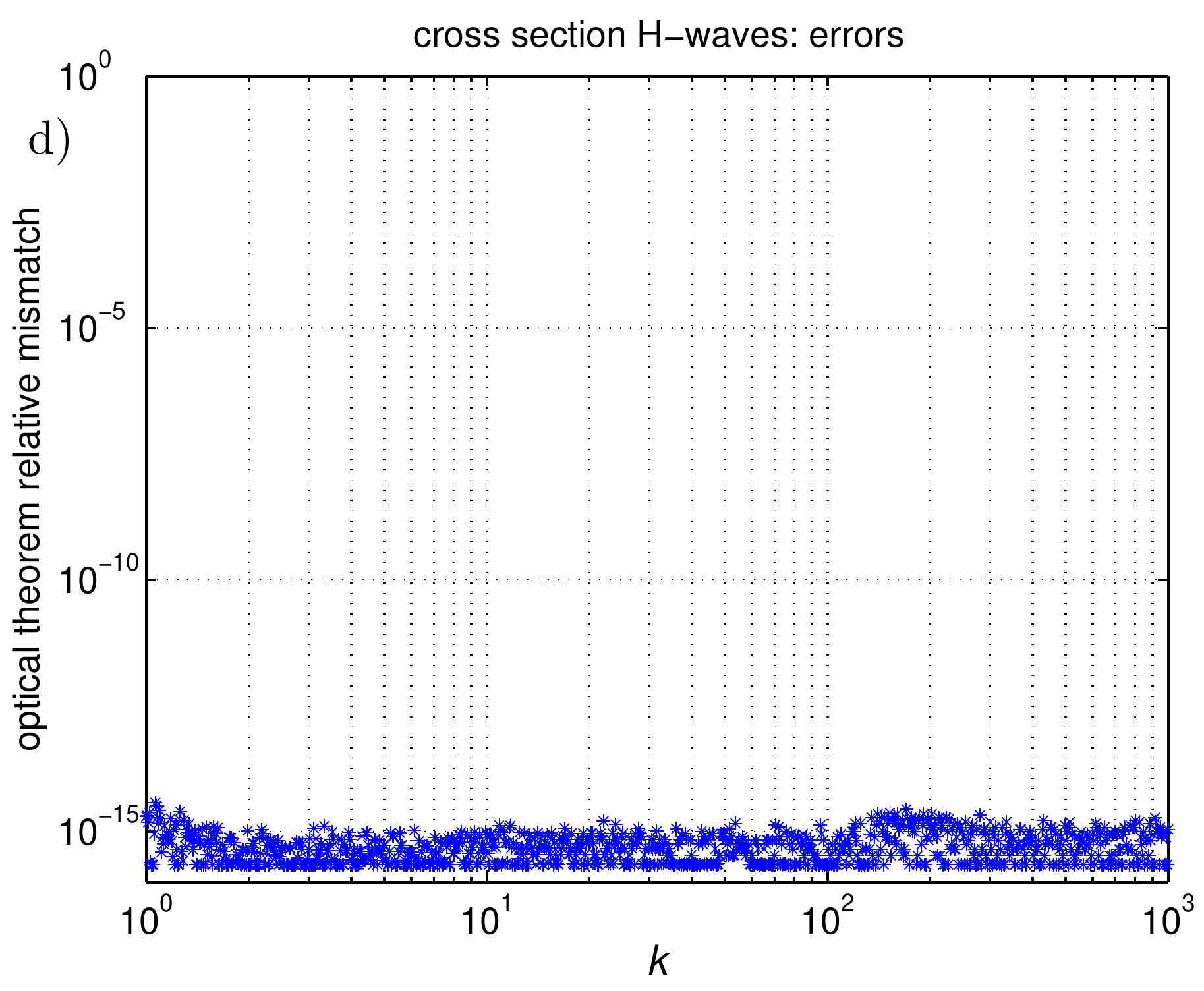}
\caption{The scattering cross sections $\sigma_\text{sca}$ for the 
  E-wave, a), and H-wave, c), calculated by the optical
  theorem~\eqref{cross2} and the relative error, b) and d), compared
  to the values from equation \eqref{cross1}}
\label{fig:cross}
\end{figure}

\subsection{Scattering cross section and optical theorem}
\label{sec:cross}

In two dimensions the scattering cross section reads
\begin{equation}\label{cross1}
\sigma_\text{sca}=\dfrac{P_\text{sca}}{\vec S_\text{inc}\cdot\hat{\vec y}}=\Re\left\{\dfrac{\iu}{\omega}\int_{\Gamma_{\text{circ}}} U_\text{sca}(\vec r')\frac{\partial U_\text{sca}^*(\vec r')}{\partial \nu_{r'}}\text{d}\ell'\right\},
\end{equation}
where $P_\text{sca}$ is the scattered power per unit length, $\vec
S_\text{inc}\cdot\hat{\vec y}$ is the $y-$component of the Poynting
vector of the incident field, i.e.~the incident power density, the
boundary $\Gamma_{\text{circ}}$ is a closed curve that circumscribes
the boundary $\Gamma$, and the star denotes complex conjugation.  The
expression holds for both E- and H-waves. In a numerical experiment
with the cylinder of \eqref{eq:gamma} we let $\Gamma_{\text{circ}}$ be
a circle of radius $0.55$ and with center at ${\vec r}=(0.5,0)$. Since
the diameter of the scatterer is $d=1$, the smallest distance between
the $\Gamma$ and $\Gamma_\text{circ}$ is $0.05$ and it occurs at the
corner vertex and at a point opposite to the corner vertex. For
evaluation points ${\vec r}'$ so close to the boundary, the field
$U_\text{sca}(\vec r')$ and its normal derivative are in general hard
to evaluate. But, as we have already seen in Section~\ref{sec:near},
the RCIP method and the high-order analytic quadrature outlined in
Section \ref{sec:discretization} should allow for high accuracy.

By utilizing the optical theorem we get an alternative expression for
the scattering cross section
\begin{equation}\label{cross2}
\sigma_\text{sca}=-\lim\limits_{y\rightarrow\infty}\Re\left\{\dfrac{4}{\omega}U_\text{sca}(0,y)\sqrt{\dfrac{\pi\omega y}{2}}e^{-\iu(\omega y-\pi/4)}\right\}
\end{equation}
which should be even simpler to evaluate than \eqref{cross1} since it
only involves the far field. The mismatch between the scattering cross
sections computed via \eqref{cross1} and via \eqref{cross2} can be
used as an error estimate for both expressions. The cross sections for
the E-waves along with such error estimates are given in Figure
\ref{fig:cross} a) and b) and the corresponding data for the H-waves
are given in Figures \ref{fig:cross} c) and d). The mismatch error is
on the order of $10^{-15}$. The cross sections in Figures
\ref{fig:cross} a) and c) show the well known behaviors for large and
small values of $k$.

\section{Conclusions}

We have shown how the basic problems of scattering of E- and H-waves
from perfectly conducting cylinders with corners can be solved
numerically to high accuracy on a mesh that on a global level is {\it
  not} refined close to corner vertices. We give examples where the
scattered electric and magnetic fields from a cylinder with one corner
and with a diameter of up to 160 wavelengths is obtained with 14
digits of accuracy almost everywhere outside the cylinder. This success is
achieved by
\begin{enumerate}
\item choosing a suitable integral representation of the scattered
  field in terms of an unknown layer density
\item formulating the scattering problem as a Fredholm second kind
  integral equation with operators that are compact away from the
  corners
\item discretizing using a Nystr{\"o}m scheme and a mix of composite
  Gauss--Legendre quadrature and high-order analytic product rules
\item modifying the discretized integral equation so that the new
  unknown, a transformed layer density, is piecewise smooth
\item solving the resulting well-conditioned linear system iteratively
  for the transformed layer density
\item partially reconstructing the original layer density from the
  transformed layer density
\item evaluating the scattered field from a discretization of its
  integral representation which, again, relies on a mix of composite
  Gauss--Legendre quadrature and high-order analytic product rules
\end{enumerate}
While some steps in this scheme are standard, step 4, 6, and 7 are
unique to the recently developed RCIP method. Conceptually, step 4 and
5 correspond to applying a fast direct solver~\cite{Kong2011} locally
to regions with troublesome geometry and then applying a global
iterative method. This gives us many of the advantages of fast direct
methods, for example the ability to deal with certain classes of
operators whose spectra make them unsuitable for iterative methods. In
addition, this approach is typically much faster than using only a
fast direct solver.

Our numerical scheme can be extended to related problems of importance
in e.g.~band-gap structures, axially symmetric cavities for accelerators, and remote sensing of underground objects.  Thus we can extend the method to scattering from homogeneous dielectric
cylinders, scattering from multiple cylinders, scattering from
cylinders in layered structures (c.f.~\cite{Karlsson1982}), scattering
of plane waves at oblique angles from cylinders, and scattering from
axially symmetric three-dimensional geometries. Some of these problems
will be addressed in forthcoming papers.

\section*{Acknowledgment}
This work was supported in part by the Swedish Research Council under
contract 621-2011-5516.

%A challenge is to generalize the method to three-dimensional problems.
%Even though the general technique used in this paper for the
%two-dimensional problem is applicable in three dimensions the
%generalization is not straightforward. Despite this, an efficient
%handling of scattering from objects with corners is important and a
%generalization of the method in this paper to three dimensions may
%lead to more efficient solvers for three-dimensional problems.

%\referencelist
\bibliography{REFERENSER}
\bibliographystyle{osajnl}
\end{document}